\documentclass[10pt]{article}

\usepackage[dvipsnames]{xcolor}

\usepackage{float}
\usepackage{graphicx}
\usepackage{tikz-inet}
\usepackage{amsfonts}
\usepackage{amsmath}
\usepackage{amsthm}
\usepackage{enumitem}

\allowdisplaybreaks

\usetikzlibrary{arrows.meta, positioning, calc}

\tikzset{every picture/.style={line cap=round, line join=round, line width=0.8pt, >={Latex[scale=1]}}}

\usepackage{duckuments}

\usepackage[justification=centering]{caption}

\usetikzlibrary{patterns}

\usepackage[precision=2, unit=mm]{lengthconvert}

\usepackage{multicol}
\usepackage{blindtext}

\usepackage[a4paper, total={16.6cm}, left=2cm, right=2cm, top=2cm, bottom=2cm]{geometry}

\usepackage{setspace}

\usepackage{titlesec}
\titleformat{\section}
  {\normalfont\large\bfseries}{\thesection.}{0.2cm}{}

\titleformat{\subsection}
  {\normalfont\large\bfseries}{\thesubsection.}{0.2cm}{}

\titleformat{\subsubsection}
  {\normalfont\normalsize\bfseries}{\thesubsubsection.}{0.2cm}{}

\titlespacing\section{0pt}{10pt}{4pt}
\titlespacing\subsection{0pt}{8pt}{4pt}
\titlespacing\subsubsection{0pt}{6pt}{3pt}

\usepackage{hyperref}
\hypersetup{
  pdftitle={Delta-Nets: Interaction-Based System for Optimal Parallel Lambda-Reduction},
  pdfauthor={Daniel Augusto Rizzi Salvadori},
  colorlinks=true,
  linkcolor=.,
  anchorcolor=.,
  citecolor=.,
  filecolor=.,
  urlcolor=.,
}

\newcommand{\fan}[4]{
  \begin{scope}[shift={(#2)}, yscale={#3}]
    \coordinate (#1-left) at (-0.7,0.4);
    \coordinate (#1-right) at (0.7,0.4);
    \coordinate (#1-tip) at (0,-0.8);
    \ifthenelse{\equal{#4}{true}}{
      \draw[line width=1.4pt]  (-1,0.4) -- (1,0.4) -- (0,-0.8) -- cycle;
    }{false};
  \end{scope}
}

\newcommand{\eraser}[4]{
  \begin{scope}[shift={(#2)}, yscale={#3}]
    \def\r{0.3}
    \def\diag{1.41421356237*\r/2}
    \coordinate (#1-tip) at (0,-\r);
    \ifthenelse{\equal{#4}{true}}{
      \draw[line width=1.4pt] (0,0) circle (\r);
      \draw[line width=1.4pt] (\diag,\diag) --  (-\diag,-\diag);
      \draw[line width=1.4pt] (\diag,-\diag) --  (-\diag,\diag);
    }{false};
  \end{scope}
}

\newcommand{\replicator}[5]{
  \begin{scope}[shift={(#2)}, xscale={#3}, yscale={#4}]
    \def\boxsize{0.9}
    \coordinate (#1-tip) at (0,-1);
    \coordinate (#1-d0) at (-2+\boxsize/2,0.4+\boxsize/2);
    \coordinate (#1-d1) at (-2+\boxsize+\boxsize/2,0.4+\boxsize/2);
    \coordinate (#1-dn) at (2-\boxsize/2,0.4+\boxsize/2);
    \coordinate (#1-d0w) at (-2+\boxsize/2,0.4+\boxsize);
    \coordinate (#1-d1w) at (-2+\boxsize+\boxsize/2,0.4+\boxsize);
    \coordinate (#1-dnw) at (2-\boxsize/2,0.4+\boxsize);
    \coordinate (#1-dots) at ({-2+2*\boxsize+(4-3*\boxsize)/2}, 0.4+\boxsize/2);
    \ifthenelse{\equal{#5}{true}}{
      \draw[line width=1.4pt]  (-2,0.4) -- (2,0.4) --  (0,-1) -- cycle;
      \draw[line width=0.8pt]  (-2,0.4) -- (-2,0.4+\boxsize) -- (-2+\boxsize,0.4+\boxsize) -- (-2+\boxsize,0.4);
      \draw[line width=0.8pt]  (-2+\boxsize,0.4+\boxsize) -- (-2+2*\boxsize,0.4+\boxsize) -- (-2+2*\boxsize,0.4);
      \draw[line width=0.8pt]  (2,0.4) -- (2,0.4+\boxsize) -- (2-\boxsize,0.4+\boxsize) -- (2-\boxsize,0.4);
      \node at (#1-dots) {$\ldots$};
    }{false};
  \end{scope}
}

\begin{document}
\begin{spacing}{1.125}

\title{\vspace{-1.125cm}\bf{\mbox{$\Delta$-Nets}: Interaction-Based System for\\Optimal Parallel \mbox{$\lambda$-Reduction}}}
\author{Daniel Augusto Rizzi Salvadori}
\date{\vspace{12pt}}
\maketitle

\begin{multicols}{2}

\section*{Abstract}

I present a model of universal parallel computation called \mbox{$\Delta$-Nets}, and a method to translate \mbox{$\lambda$-terms} into \mbox{$\Delta$-nets} and back.
Together, the model and the method constitute an algorithm for optimal parallel \mbox{$\lambda$-reduction}, solving the longstanding enigma with groundbreaking clarity.
I show that the \mbox{$\lambda$-calculus} can be understood as a projection of \mbox{$\Delta$-Nets}---one that severely restricts the structure of sharing, among other drawbacks.
Unhindered by these restrictions, the \mbox{$\Delta$-Nets} model opens the door to new parallel programming language implementations and computer architectures that are more efficient and performant than previously possible.

\vspace{1.4em}
\noindent \textbf{Interactive Demo:} \url{https://deltanets.org}
\vspace{0.4em}

\section{Introduction}

The \mbox{$\lambda$-calculi} are beautifully simple yet powerful models of universal computation. Consisting of \textit{abstractions}, \textit{variables}, and \textit{applications}, \mbox{$\lambda$-terms} can express any computable function \cite{Chu36, Tur36, Tur37, Chu41}. In addition to being central pillars in computation theory, the \mbox{$\lambda$-calculi} also constitute practical frameworks underpinning all functional programming languages.
Four \mbox{$\lambda$-calculi} are pertinent to this paper---the three substructure \mbox{$\lambda$-calculi} \cite{Jac93}, and the full \mbox{$\lambda$-calculus} \cite{Bar84}:

\begin{itemize}
\itemsep0em
\item \textbf{\mbox{$\lambda L$-calculus}}: the \textit{linear} \mbox{$\lambda$-calculus}, in which every bound variable occurs exactly once.
\item \textbf{\mbox{$\lambda A$-calculus}}: the \textit{affine} \mbox{$\lambda$-calculus}, in which every bound variable occurs either once or not at all.
\item \textbf{\mbox{$\lambda I$-calculus}}\footnote{the original \mbox{$\lambda$-calculus} defined by Church \cite{Chu36,Chu41}.}:
the \textit{relevant} \mbox{$\lambda$-calculus}, in which every bound variable occurs at least once.
\item \textbf{\mbox{$\lambda K$-calculus}}: the \textit{full} \mbox{$\lambda$-calculus} in which bound variables can occur any number of times.
\end{itemize}

\begin{figure}[H]
  \vspace{-5pt}
  \centering
  \begin{tikzpicture}[scale=0.58]
    \Large

    \node (L) at (0, 2) {$\lambda L$};
    \node (A) at (-2, 0) {$\lambda A$};
    \node (I) at (2, 0) {$\lambda I$};
    \node (K) at (0, -2) {$\lambda K$};

    \coordinate (A_topright) at ($(A) + (45:1cm)$);
    \coordinate (A_bottomleft) at ($(A) + (45+180:1cm)$);
    \coordinate (I_topleft) at ($(I) + (45+90:1cm)$);
    \coordinate (I_bottomright) at ($(I) + (45+180+90:1cm)$);
    \coordinate (K_topright) at ($(K) + (45:1cm)$);
    \coordinate (K_topleft) at ($(K) + (45+90:1cm)$);
    \coordinate (K_bottomright) at ($(K) + (-45:1cm)$);
    \coordinate (K_bottomleft) at ($(K) + (-45-90:1cm)$);

    \draw[->] (L) -- (A);
    \draw[->] (L) -- (I);
    \draw[->] (L) -- (K);
    \draw[->] (A) -- (K);
    \draw[->] (I) -- (K);

    \draw[dotted, line cap=butt, line join=butt] (K_topright) arc[start angle=45, end angle=-180+45, radius=1cm] -- (A_bottomleft) arc[start angle=180+45, end angle=45, radius=1cm] -- cycle;
    \draw[dotted, line cap=butt, line join=butt] (K_topleft) arc[start angle=135, end angle=135+180, radius=1cm] -- (I_bottomright) arc[start angle=-45, end angle=135, radius=1cm] -- cycle;

    \normalsize
    \node at (-3, -2) {erasure};
    \node at (3, -2) {sharing};
  \end{tikzpicture}

  \caption{The relationships between the four \mbox{$\lambda$-calculi} in terms of erasure and sharing.}
  \label{fig:calculi}
\end{figure}
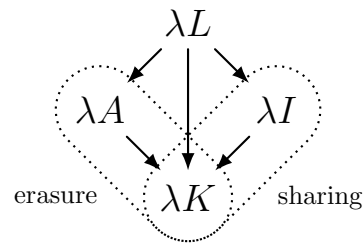

The \mbox{$\lambda L$-calculus} can be regarded as a cornerstone which can be extended in one of three ways: with \textit{erasure} (analogous to \textit{weakening} in logic), resulting in the \mbox{$\lambda A$-calculus}; with \textit{sharing} (analogous to \textit{contraction} in logic), resulting in the \mbox{$\lambda I$-calculus}; or with \textit{both erasure and sharing}, resulting in the full \mbox{$\lambda K$-calculus}. Additionally, the \mbox{$\lambda K$-calculus} can also be obtained by extending the \mbox{$\lambda A$-calculus} with sharing, or the \mbox{$\lambda I$-calculus} with erasure. These relationships are illustrated in \mbox{Figure \ref{fig:calculi}}.

\vspace{0.2cm}
\vspace{0.2cm}

In the \mbox{$\lambda$-calculi}, a reduction strategy is \textit{optimal} if and only if it reaches the normal form (if it exists) without perfoming any unnecessary reduction steps \cite{Lev78, Lev80}. There are two types of unnecessary reductions:

\indent1. Reduction of a later-discarded subexpression.\\
\indent2. Reduction of a duplicated subexpression.

The first type can only occur with erasure, i.e., when some abstractions do not make use of their bound variables. Whenever such an abstraction is applied, the argument is discarded, and any reductions previously performed in the argument subexpression are rendered unnecessary. Naturally, this cannot happen in the \mbox{$\lambda L$-calculus} nor in the \mbox{$\lambda I$-calculus}. This first type of unnecessary reduction can be entirely avoided in the \mbox{$\lambda A$-calculus} and in the \mbox{$\lambda K$-calculus} by adhering to some reduction orders, including normal order reduction \cite{Bar87}.

The second type of unnecessary reduction can only occur with sharing, i.e., when some abstractions' bound variables occur multiple times. Whenever such an abstraction is applied, the argument is duplicated, and any reducible expressions in the argument subexpression are duplicated with it. Naturally, this cannot happen in the \mbox{$\lambda L$-calculus} nor in the \mbox{$\lambda A$-calculus}.
Lévy has shown that there are \mbox{$\lambda$-terms} with sharing for which no reduction order is optimal \cite{Lev78, Lev80}. The \mbox{$\lambda$-calculus}, as a sequential substitution machine, is therefore inadequate to express optimal reduction for all \mbox{$\lambda$-terms}. Does a more fundamental model of computation exist which is able to express optimal reduction for all \mbox{$\lambda$-terms}?

In order to avoid sharing-related unnecessary reductions in the \mbox{$\lambda I$-} and \mbox{$\lambda K$-calculi}, it is useful to represent \mbox{$\lambda$-terms} as graphs.
Identical subexpressions can then be represented by the same shared subgraph, which only needs to be reduced once. The process of reduction is then expressed as a sequence of graph operations---a technique known as graph reduction \cite{Wad71}.
In the \mbox{$\lambda$-calculi}, applying a function destructively modifies its body. This presents a challenge in graph reduction when a function is shared, since it may be applied any number of times, each with a different argument. A simple solution is to duplicate the shared function's entire subgraph before applying it. This solution, however, leads to the second type of unnecessary reduction because it duplicates reducible expressions. It's possible to mitigate the number of duplicated reducible expressions, and thus of unnecessary reductions, by sharing, instead of duplicating, the function's \textit{maximal free subgraphs}---the largest subgraphs in the function's body that don't make use of the function's bound variable \cite{Wad71}.
An ordering can also be imposed such that all reducible expressions inside a function are first reduced, and only then is the function duplicated (while sharing its maximal free subgraphs). However, a critical problem remains: this procedure never terminates in cyclic graphs, and cyclic graphs can arise from non-cyclic ones through regular reduction \cite{Wad71}. In \cite{Wad71} this is resolved by duplicating subgraphs whenever necessary to avoid cycles.

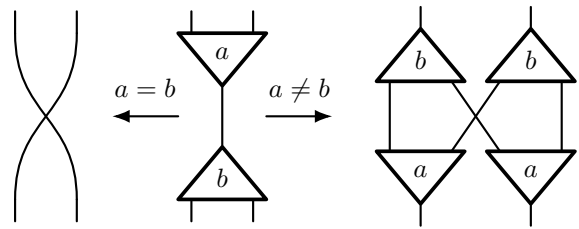
\begin{figure}[H]
  \centering
  \begin{tikzpicture}[scale=0.58]

    \begin{scope}

      \fan{fan1}{0,0}{1}{false};
      \fan{fan2}{0,-3}{-1}{false};

      \draw (fan1-left) .. controls ++(0, -1.9) and ++(0, 1.9) .. (fan2-right);
      \draw (fan1-right) .. controls ++(0, -1.9) and ++(0, 1.9) .. (fan2-left);

      \draw (fan1-left) -- ++(0, 0.5);
      \draw (fan1-right) -- ++(0, 0.5);
      \draw (fan2-left) -- ++(0, -0.5);
      \draw (fan2-right) -- ++(0, -0.5);
    \end{scope}

    \draw[->] (3, -1.5) -- (1.5, -1.5) node[midway,above,yshift=2.5mm,anchor=base] {$a = b$};

    \begin{scope}[xshift=4cm]

      \fan{fan1}{0,0}{1}{true};
      \node at (0,-0.05) {$a$};
      \fan{fan2}{0,-3}{-1}{true};
      \node at (0,-2.9) {$b$};

      \draw (fan1-tip) -- (fan2-tip);
      \draw (fan1-left) -- ++(0, 0.5);
      \draw (fan1-right) -- ++(0, 0.5);
      \draw (fan2-left) -- ++(0, -0.5);
      \draw (fan2-right) -- ++(0, -0.5);
    \end{scope}

    \draw[->] (5, -1.5) -- (6.5, -1.5) node[midway,above,yshift=2.5mm,anchor=base] {$a \neq b$};

    \begin{scope}[xshift=8.5cm]

      \fan{fan1}{0,-0.3}{-1}{true};
      \node at (0,-0.2) {$b$};
      \fan{fan2}{2.5,-0.3}{-1}{true};
      \node at (2.5,-0.2) {$b$};
      \fan{fan3}{0,-2.7}{1}{true};
      \node at (0,-2.75) {$a$};
      \fan{fan4}{2.5,-2.7}{1}{true};
      \node at (2.5,-2.75) {$a$};

      \draw (fan1-tip) -- ++(0, 0.5);
      \draw (fan2-tip) -- ++(0, 0.5);
      \draw (fan3-tip) -- ++(0, -0.5);
      \draw (fan4-tip) -- ++(0, -0.5);
      \draw (fan1-left) -- (fan3-left);
      \draw (fan1-right) -- (fan4-left);
      \draw (fan2-left) -- (fan3-right);
      \draw (fan2-right) -- (fan4-right);
    \end{scope}

  \end{tikzpicture}
  \caption{Interacting fans annihilate one another if
  they are equal, otherwise they duplicate one another.}
  \label{fig:fanrule}
\end{figure}

The first algorithms for \mbox{$\lambda$-calculi} reduction which don't perform any unnecessary \mbox{$\beta$-reductions} were proposed in \cite{Lam89} and \cite{Kat90}. The common core idea introduced was that of interior sharing of subgraphs. Interior sharing enables the incremental duplication of shared functions, which, in turn, fully prevents sharing-related unnecessary \mbox{$\beta$-reductions}. The challenge of interior sharing lies in the management, throughout reduction, of multiple simultaneous sharing contexts, each of which can fully or partially overlap any number of others. Moreover, sharing contexts can be recursive---a shared term can be referenced from inside itself any number of times. In \cite{Lam89}, interior sharing is accomplished through the use of explicit fan-in and fan-out nodes. Remarkably, the complex challenge then boils down to a simple question: when a particular fan-in meets a particular fan-out, should they annihilate one another or duplicate one another? In \cite{Lam89}, this is solved by associating a non-negative integer with each fan and introducing three delimiter node types to regulate fan numbers through graph-rewriting rules. When two fans meet, they are annihilated if and only if their numbers are equal, and they duplicate one another otherwise (\mbox{Figure \ref{fig:fanrule}}). The delimiters realize a notion of enclosure around fans, ultimately ensuring that only fans belonging to the same enclosure annihilate one another.

\vspace{0.2cm}
\vspace{0.2cm}

Concurrently, a graphical meta-model of parallel computation called \textit{interaction nets} was introduced in \cite{Laf90}. It was later refined and expanded in \cite{Laf97}, once its significance was understood better.
Models adhering to this framework are called \textit{interaction systems}. An interaction system is specified by a set of agent types and a set of interaction rules---local graph rewriting rules that operate on pairs of connected agents. Agents are nodes that have one \textit{principal} port and any number of \textit{auxiliary} ports, including zero. When two agents are connected via their principal ports they form an \textit{active pair}, and an interaction rule can then be applied to that pair. Since each agent can only be part of a single active pair at a time, interaction systems possess a one-step diamond property, which I denote as \textit{perfect confluence}\footnote{Some authors denote this as \textit{strong confluence}, but strong confluence has historically denoted a weaker property which is not one-step. I propose using the term \textit{perfect confluence} for the one-step variant to avoid confusion.}.
An extraordinary consequence of perfect confluence is that every normalizing interaction order produces the same result in the same number of interactions.
Additionally, since interaction rules are local, they can be applied simultaneously without synchronization. These properties make interaction systems highly suitable for expressing optimal parallel algorithms.

An interaction system for \mbox{$\lambda$-calculi} reduction based on \cite{Lam89}, along with a translation method from and to \mbox{$\lambda$-terms}, was proposed in \cite{GAL92a, GAL92b}. A similar interaction system, combined with a different translation method, was presented in \cite{AG98}.
These interaction systems employ indexed agents called \textit{brackets} and \textit{croissants}, which increment and decrement, respectively, the index of higher-index agents they interact with.
As explained in detail in \cite{AG98}, there are many ways to translate \mbox{$\lambda$-terms} into interaction nets using brackets and croissants.
These agents effectively delimit sharing scopes, and, as such, are also referred to as \textit{delimiters}.
A comprehensive analysis and comparison of the computational efficiency of these algorithms, including the original \cite{Lam89},
was supplied in \cite{LM99}. Unfortunately, all these algorithms have frustrating characteristics that profoundly undermine reduction performance. Critically, delimiters accumulate during reduction until delimiter interactions completely overwhelm fan interactions. Additionally, delimiters are often present in nets associated with \mbox{$\lambda$-terms} that have no sharing, serving no purpose there.

A more recent interaction system for \mbox{$\lambda$-calculi} reduction, called \textit{Lambdascope}, was presented in \cite{OL04}. It uses five agent types: an \textit{abstractor}, an \textit{applicator}, a \textit{duplicator} (an indexed fan), an \textit{eraser}, and an indexed delimiter. In Lambdascope, applying an abstraction effectively spawns two zero-indexed delimiters. This process, along with other interaction rules, preserves the scope of each abstraction after it has been applied. This \textit{scope invariant}, in turn, ensures interacting duplicators annihilate and commute when appropriate. Delimiters move outward, expanding each scope as much as possible. Sibling scopes eventually coalesce, which helps reduce the number of total scopes and thus the number of total delimiters. However, siblingless scopes and their delimiters are preserved perpetually. As a result, Lambdascope also suffers from a devastating accumulation of delimiters. Additionally, since there's no mechanism to decrement indexes, they grow without bound as reduction progresses.

For example, when reducing the non-normalizing $\lambda$-term $(\lambda x.x\,x)(\lambda y.y\,y)$ with these algorithms, every iteration creates new delimiters. Each new delimiter takes up additional memory (space), and leads to additional unnecessary interactions (time).
Additionally, the existing algorithms fail to establish a global reduction order, which is unfortunately critical to ensure that all nets associated with normalizing \mbox{$\lambda$-terms} normalize.
Not only are the existing algorithms unsatisfactory from a theoretical perspective, their inefficiencies preclude them from being used at the core of programming language implementations.

In this paper, I present a model of universal parallel computation called {$\Delta$-Nets}, and a method to translate \mbox{$\lambda$-terms} into \mbox{$\Delta$-nets} and back. Together, the system and translation method constitute an algorithm for optimal parallel \mbox{$\lambda$-reduction}.
As an interaction system, the core of the model is perfectly confluent: every normalizing reduction order produces the same result in the same number of steps.
The core model is then necessarily extended with non-interaction \emph{canonicalization} rules, which depart from the interaction paradigm.
Along with a global reduction order, canonicalizations ensure Church--Rosser confluence and optimality in nonlinear systems.
The \mbox{$\Delta$-Nets} algorithm solves the longstanding enigma of optimal \mbox{$\lambda$-calculi} reduction with groundbreaking clarity.
Instead of making use of delimiters, sharing is expressed through a single agent type which allows any number of auxiliary ports, called a \textit{replicator}.
Each instance of a replicator incorporates information that in previous models was spread across multiple agents, such as indexed fans and delimiters. This consolidation of information enables simplifications that were previously unfeasible, and leads to constant memory usage in the reduction of $(\lambda x.x\,x)(\lambda y.y\,y)$, for example.
%
Finally, I show that the \mbox{$\lambda$-calculus} can be understood as a projection of \mbox{$\Delta$-Nets}. The additional degrees of freedom in \mbox{$\Delta$-Nets} allow it to realize optimal reduction in the manner envisioned by Lévy, i.e., no reduction operation is applied which is rendered unneccessary later, and no reduction operation which is necessary is applied more than once.

\section{Core Interaction System}

At its core, a \mbox{$\Delta$-net} is an interaction net with three agent types: a \textit{fan}, an \textit{eraser}, and a \textit{replicator}.

\vspace{0.2cm}
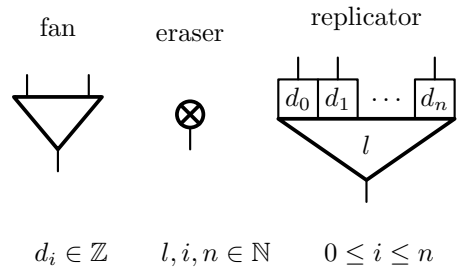
\begin{figure}[H]
  \centering
  \begin{tikzpicture}[scale=0.58]
    \begin{scope}
      \node at (0, 2) {fan};
      \fan{fan1}{0,0}{1}{true};
      \draw (fan1-left) -- ++(0, 0.5);
      \draw (fan1-right) -- ++(0, 0.5);
      \draw (fan1-tip) -- ++(0, -0.5);
    \end{scope}

    \begin{scope}[xshift=3cm]
      \node at (0, 1.8) {eraser};
      \eraser{era1}{0,0}{1}{true};
      \draw (era1-tip) -- ++(0, -0.5);
    \end{scope}

    \begin{scope}[xshift=7cm, yshift=-0.2cm]
      \node at (0, 2.4) {replicator};
      \replicator{rep1}{0,-0.3}{1}{1}{true};

      \draw (rep1-tip) -- ++(0, -0.5);
      \draw (rep1-d0w) -- ++(0, 0.5);
      \draw (rep1-d1w) -- ++(0, 0.5);
      \draw (rep1-dnw) -- ++(0, 0.5);
      \node at (rep1-d0) {$d_0$};
      \node at (rep1-d1) {$d_1$};
      \node at (rep1-dn) {$d_n$};
      \node[anchor=base] at (0,-0.7) {$l$};


    \end{scope}

    \begin{scope}
      \node[anchor=center] at (4, -3.2) {$d_i \in \mathbb{Z}\qquad l,i,n \in \mathbb{N}\qquad 0 \le i \le n$};
    \end{scope}

  \end{tikzpicture}
  \caption{The three $\Delta$-Nets agent types.}
  \label{fig:lnetsagents}
\end{figure}
\vspace{0.2cm}

Fans have two auxiliary ports, and erasers have none. Each replicator can have a different natural number of auxiliary ports,
and each of those ports has an associated integer called a \textit{level delta}.
Additionally, each replicator has an associated non-negative integer called a \textit{level}.
As a point of reference, an agent's auxiliary ports are ordered in a clockwise orientation, i.e., from left to right when the principal port points down.

\begin{figure*}[!b]
  \centering
  \begin{tikzpicture}[scale=0.58]

    \begin{scope}[xshift=-0.25cm]
      \begin{scope}

        \eraser{era1}{0,-0.5}{1}{true};
        \eraser{era2}{0,-3}{-1}{true};

        \draw[dotted] (-1.25, 0.8) -- (1.25, 0.8) -- (1.25, -3.3-1) -- (-1.25, -3.3-1) -- cycle;

        \draw (era1-tip) -- (era2-tip);
      \end{scope}

      \draw[->] (1.25, -1.75) -- (2.75, -1.75);

      \begin{scope}[xshift=4cm]
        \draw[dotted] (-1.25, 0.8) -- (1.25, 0.8) -- (1.25, -3.3-1) -- (-1.25, -3.3-1) -- cycle;
      \end{scope}
    \end{scope}

    \begin{scope}[xshift=8.25cm]
      \begin{scope}

        \eraser{era1}{0,-0.5}{1}{true};
        \fan{fan1}{0,-4.3 + 0.4 + 1}{-1}{true};

        \draw[dotted] (-1.5, 0.8) -- (1.5, 0.8) -- (1.5, -3.3-1) -- (-1.5, -3.3-1) -- cycle;

        \draw (era1-tip) -- (fan1-tip);
        \draw (fan1-left) -- ++(0, -1.75);
        \draw (fan1-right) -- ++(0, -1.75);
      \end{scope}

      \draw[->] (1.5, -1.75) -- (3, -1.75);

      \begin{scope}[xshift=4.5cm]

        \eraser{era1}{-0.7,-3.25}{1}{true};
        \eraser{era2}{0.7,-3.25}{1}{true};
        \fan{fan1}{0,-3}{-1}{false};

        \draw[dotted] (-1.5, 0.8) -- (1.5, 0.8) -- (1.5, -3.3-1) -- (-1.5, -3.3-1) -- cycle;

        \draw (era1-tip) -- ++(0, -1.5);
        \draw (era2-tip) -- ++(0, -1.5);
      \end{scope}
    \end{scope}

    \begin{scope}[xshift=18.5cm]
      \begin{scope}
        \replicator{rep1}{0,-1}{1}{1}{true};
        \eraser{era1}{0,-3.3}{-1}{true};

        \draw[dotted] (-2.5, 0.8) -- (2.5, 0.8) -- (2.5,-3.3-1) -- (-2.5,-3.3-1) -- cycle;

        \draw (rep1-tip) -- (era1-tip);
        \draw (rep1-d0w) -- ++(0,1.25);
        \draw (rep1-d1w) -- ++(0,1.25);
        \draw (rep1-dnw) -- ++(0,1.25);
        \node at (rep1-d0) {$d_0$};
        \node at (rep1-d1) {$d_1$};
        \node at (rep1-dn) {$d_n$};
        \node[anchor=base] at (0,-1.4) {$l$};

        \node at (rep1-dots |-, 1.25) {$\ldots$};
      \end{scope}

      \draw[->] (2.5, -1.75) -- (4, -1.75);

      \begin{scope}[xshift=6.5cm]
        \replicator{rep1}{0,-1}{1}{1}{false};
        \eraser{era1}{rep1-d0}{-1}{true};
        \eraser{era2}{rep1-d1}{-1}{true};
        \eraser{era3}{rep1-dn}{-1}{true};

        \draw[dotted] (-2.5, 0.8) -- (2.5, 0.8) -- (2.5,-3.3-1) -- (-2.5,-3.3-1) -- cycle;

        \draw (era1-tip) -- ++(0, 1.35);
        \draw (era2-tip) -- ++(0, 1.35);
        \draw (era3-tip) -- ++(0, 1.35);

        \node at (rep1-dots |-, 1.25) {$\ldots$};
        \node at (rep1-dots) {$\ldots$};
      \end{scope}

    \end{scope}


    \begin{scope}[xshift=0.5cm, yshift=-7.8cm]
      \begin{scope}

        \fan{fan1}{0,-0.1}{1}{true};
        \fan{fan2}{0,-4.1}{-1}{true};

        \draw[dotted] (-2, 0.8 + 0.5) -- (2, 0.8 + 0.5) -- (2, -4-1-0.5) -- (-2, -4-1-0.5) -- cycle;

        \draw (fan1-tip) -- (fan2-tip);
        \draw (fan1-left) -- ++(0, 1.75);
        \draw (fan1-right) -- ++(0, 1.75);
        \draw (fan2-left) -- ++(0, -1.75);
        \draw (fan2-right) -- ++(0, -1.75);
      \end{scope}

      \draw[->] (2, -2.1) -- (4-0.125, -2.1);

      \begin{scope}[xshift=1cm-0.125cm, xshift=5cm]

        \fan{fan1}{0,-0.1}{1}{false};
        \fan{fan2}{0,-4.1}{-1}{false};

        \draw[dotted] (-2, 0.8 + 0.5) -- (2, 0.8 + 0.5) -- (2, -4-1-0.5) -- (-2, -4-1-0.5) -- cycle;

        \draw (fan1-left) .. controls ++(0, -3) and ++(0, 3) .. (fan2-right);
        \draw (fan1-right) .. controls ++(0, -3) and ++(0, 3) .. (fan2-left);

        \draw (fan1-left) -- ++(0, 1.75);
        \draw (fan1-right) -- ++(0, 1.75);
        \draw (fan2-left) -- ++(0, -1.75);
        \draw (fan2-right) -- ++(0, -1.75);
      \end{scope}
    \end{scope}

    \begin{scope}[xshift=13cm+0.125cm, yshift=-7.8cm]
      \begin{scope}
        \replicator{rep1}{0,-0.5}{1}{1}{true};
        \fan{fan1}{0,-3.8-0.2-0.6+0.5}{-1}{true};

        \draw[dotted] (-2.5, 0.8 + 0.5) -- (2.5, 0.8 + 0.5) -- (2.5,-4-1-0.5) -- (-2.5,-4-1-0.5) -- cycle;

        \draw (rep1-tip) -- (fan1-tip);
        \draw (rep1-d0w) -- ++(0,1.25);
        \draw (rep1-d1w) -- ++(0,1.25);
        \draw (rep1-dnw) -- ++(0,1.25);
        \draw (fan1-left) -- ++(0,-1.75);
        \draw (fan1-right) -- ++(0,-1.75);
        \node at (rep1-d0) {$d_0$};
        \node at (rep1-d1) {$d_1$};
        \node at (rep1-dn) {$d_n$};
        \node[anchor=base] at (0,-0.9) {$l$};

        \node at (rep1-dots |-, 1.75) {$\ldots$};
      \end{scope}

      \draw[->] (2.5, -2.1) -- (4.5-0.125, -2.1);

      \begin{scope}[xshift=7cm-0.125cm]
        \fan{fan1}{-0.85,0}{-1}{true};
        \fan{fan2}{-0.85 + 3,0}{-1}{true};
        \fan{fan3}{-0.85 + 6.7,0}{-1}{true};
        \replicator{rep1}{0,-4}{1}{1}{true};
        \replicator{rep2}{5,-4}{1}{1}{true};

        \draw[dotted] (-2.5, 0.8 + 0.5) -- (7.5, 0.8 + 0.5) -- (7.5,-4-1-0.5) -- (-2.5,-4-1-0.5) -- cycle;

        \draw (rep1-d0w) -- (fan1-left);
        \draw (rep1-d1w) .. controls ++(0, 1) and ++(0, -1) .. (fan2-left);
        \draw (rep1-dnw) .. controls ++(0, 1) and ++(0, -1) .. (fan3-left);

        \draw (rep2-d0w) .. controls ++(0, 1) and ++(0, -1) .. (fan1-right);
        \draw (rep2-d1w) .. controls ++(0, 1) and ++(0, -1) .. (fan2-right);
        \draw (rep2-dnw) -- (fan3-right);

        \node at (rep1-d0) {$d_0$};
        \node at (rep1-d1) {$d_1$};
        \node at (rep1-dn) {$d_n$};

        \node at (rep2-d0) {$d_0$};
        \node at (rep2-d1) {$d_1$};
        \node at (rep2-dn) {$d_n$};

        \draw (fan1-tip) -- ++(0,1.25);
        \draw (fan2-tip) -- ++(0,1.25);
        \draw (fan3-tip) -- ++(0,1.25);

        \draw (rep1-tip) -- ++(0,-1.25);
        \draw (rep2-tip) -- ++(0,-1.25);

        \node at (3.875, 0.25) {$\ldots$};
        \node at (3.875, 1.3 + 0.5) {$\ldots$};

        \node[anchor=base] at (0,-4.3) {$l$};
        \node[anchor=base] at (5,-4.4) {$l$};
      \end{scope}

    \end{scope}

    \begin{scope}[xshift=-0.5cm, yshift=-17cm]

      \begin{scope}[xshift=1.25cm]
        \replicator{rep1}{0,-1+0.7}{1}{1}{false};
        \replicator{rep2}{0,-4.2-0.5}{-1}{-1}{false};

        \draw[dotted] (-2.25, 0.8+0.7) -- (2.25, 0.8+0.7) -- (2.25,-4-1-1-0.5) -- (-2.25,-4-1-1-0.5) -- cycle;

        \draw (rep1-d0w) .. controls ++(0, -3) and ++(0, 3) .. (rep2-d0w);
        \draw (rep1-d1w) .. controls ++(0, -3) and ++(0, 3) .. (rep2-d1w);
        \draw (rep1-dnw) .. controls ++(0, -3) and ++(0, 3) .. (rep2-dnw);

        \draw (rep1-d0w) -- ++(0, 1.25);
        \draw (rep1-d1w) -- ++(0, 1.25);
        \draw (rep1-dnw) -- ++(0, 1.25);

        \draw (rep2-d0w) -- ++(0, -1.25);
        \draw (rep2-d1w) -- ++(0, -1.25);
        \draw (rep2-dnw) -- ++(0, -1.25);

        \node at (rep1-dots |-, 1.25+0.7) {$\ldots$};
        \node at (rep1-dots) {$\ldots$};
        \node at (rep2-dots |-, -6.5-0.5) {$\ldots$};
        \node at (rep2-dots) {$\ldots$};
      \end{scope}

      \draw[->] (5.5, -2.5) -- (3.5, -2.5) node[midway, above, yshift=1mm] {$l = k$};

      \begin{scope}[xshift=8cm]
        \replicator{rep1}{0,-1+0.7}{1}{1}{true};
        \replicator{rep2}{0,-4.2-0.5}{-1}{-1}{true};

        \draw[dotted] (-2.5, 0.8+0.7) -- (2.5, 0.8+0.7) -- (2.5,-4-1-1-0.5) -- (-2.5,-4-1-1-0.5) -- cycle;

        \draw (rep1-tip) -- (rep2-tip);
        \draw (rep1-d0w) -- ++(0, 1.25);
        \draw (rep1-d1w) -- ++(0, 1.25);
        \draw (rep1-dnw) -- ++(0, 1.25);

        \draw (rep2-d0w) -- ++(0, -1.25);
        \draw (rep2-d1w) -- ++(0, -1.25);
        \draw (rep2-dnw) -- ++(0, -1.25);

        \node at (rep1-d0) {$d_0$};
        \node at (rep1-d1) {$d_1$};
        \node at (rep1-dn) {$d_n$};

        \node at (rep2-d0) {$e_0$};
        \node at (rep2-d1) {$e_1$};
        \node at (rep2-dn) {$e_m$};

        \node[anchor=base] at (0,-1.4+0.7) {$l$};
        \node[anchor=center] at (0,-4.2-0.5 + 0.18) {$k$};

        \node at (rep1-dots |-, 1.25+0.7) {$\ldots$};
        \node at (rep2-dots |-, -6.5-0.5) {$\ldots$};
      \end{scope}

      \draw[->] (10.5, -2.5) -- (12.5, -2.5) node[midway, above, yshift=1mm] {$l < k$};

      \begin{scope}[xshift=16cm];
        \replicator{rep1}{-1,0}{-1}{-1}{true};
        \replicator{rep2}{4,0}{-1}{-1}{true};
        \replicator{rep3}{9.5,0}{-1}{-1}{true};
        \replicator{rep1b}{-1,-5}{1}{1}{true};
        \replicator{rep2b}{4.5,-5}{1}{1}{true};
        \replicator{rep3b}{9.5,-5}{1}{1}{true};

        \draw[dotted] (-3.5, 1 + 0.5) -- (12, 1 + 0.5) -- (12,-5-1-0.5) -- (-3.5,-5-1-0.5) -- cycle;

        \draw (rep1-d0w) .. controls ++(0, -1) and ++(0, 1) .. (rep3b-d0w);
        \draw (rep1-d1w) .. controls ++(0, -1) and ++(0, 1) .. (rep2b-d0w);
        \draw (rep1-dnw) -- (rep1b-d0w);

        \draw (rep2-d0w) .. controls ++(0, -1) and ++(0, 1) .. (rep3b-d1w);
        \draw (rep2-d1w) .. controls ++(0, -1) and ++(0, 1) .. (rep2b-d1w);
        \draw (rep2-dnw) .. controls ++(0, -1) and ++(0, 1) .. (rep1b-d1w);

        \draw (rep3-d0w) -- (rep3b-dnw);
        \draw (rep3-d1w) .. controls ++(0, -1) and ++(0, 1) .. (rep2b-dnw);
        \draw (rep3-dnw) .. controls ++(0, -1) and ++(0, 1) .. (rep1b-dnw);

        \node at (rep1-d0) {$e_0$};
        \node at (rep1-d1) {$e_1$};
        \node at (rep1-dn) {$e_m$};
        \node[anchor=center] at (-1,0.04) {$k + d_0$};

        \node at (rep2-d0) {$e_0$};
        \node at (rep2-d1) {$e_1$};
        \node at (rep2-dn) {$e_m$};
        \node[anchor=center] at (4,0.04) {$k + d_1$};

        \node at (rep3-d0) {$e_0$};
        \node at (rep3-d1) {$e_1$};
        \node at (rep3-dn) {$e_m$};
        \node[anchor=center] at (9.5,0.04) {$k + d_n$};

        \node at (rep1b-d0) {$d_0$};
        \node at (rep1b-d1) {$d_1$};
        \node at (rep1b-dn) {$d_n$};
        \node[anchor=base] at (-1,-5.4) {$l$};

        \node at (rep2b-d0) {$d_0$};
        \node at (rep2b-d1) {$d_1$};
        \node at (rep2b-dn) {$d_n$};
        \node[anchor=base] at (4.5,-5.4) {$l$};

        \node at (rep3b-d0) {$d_0$};
        \node at (rep3b-d1) {$d_1$};
        \node at (rep3b-dn) {$d_n$};
        \node[anchor=base] at (9.5,-5.4) {$l$};

        \draw (rep1-tip) -- ++(0,1.25);
        \draw (rep2-tip) -- ++(0,1.25);
        \draw (rep3-tip) -- ++(0,1.25);

        \draw (rep1b-tip) -- ++(0,-1.25);
        \draw (rep2b-tip) -- ++(0,-1.25);
        \draw (rep3b-tip) -- ++(0,-1.25);

        \node at (6.75, 0.25) {$\ldots$};
        \node at (6.75, 2) {$\ldots$};
        \node at (1.75, -5.25) {$\ldots$};
        \node at (1.75, -7) {$\ldots$};
      \end{scope}

    \end{scope}

  \end{tikzpicture}
  \caption{The core \mbox{$\Delta$-Nets} interaction rules.}
  \label{fig:lirules}
\end{figure*}
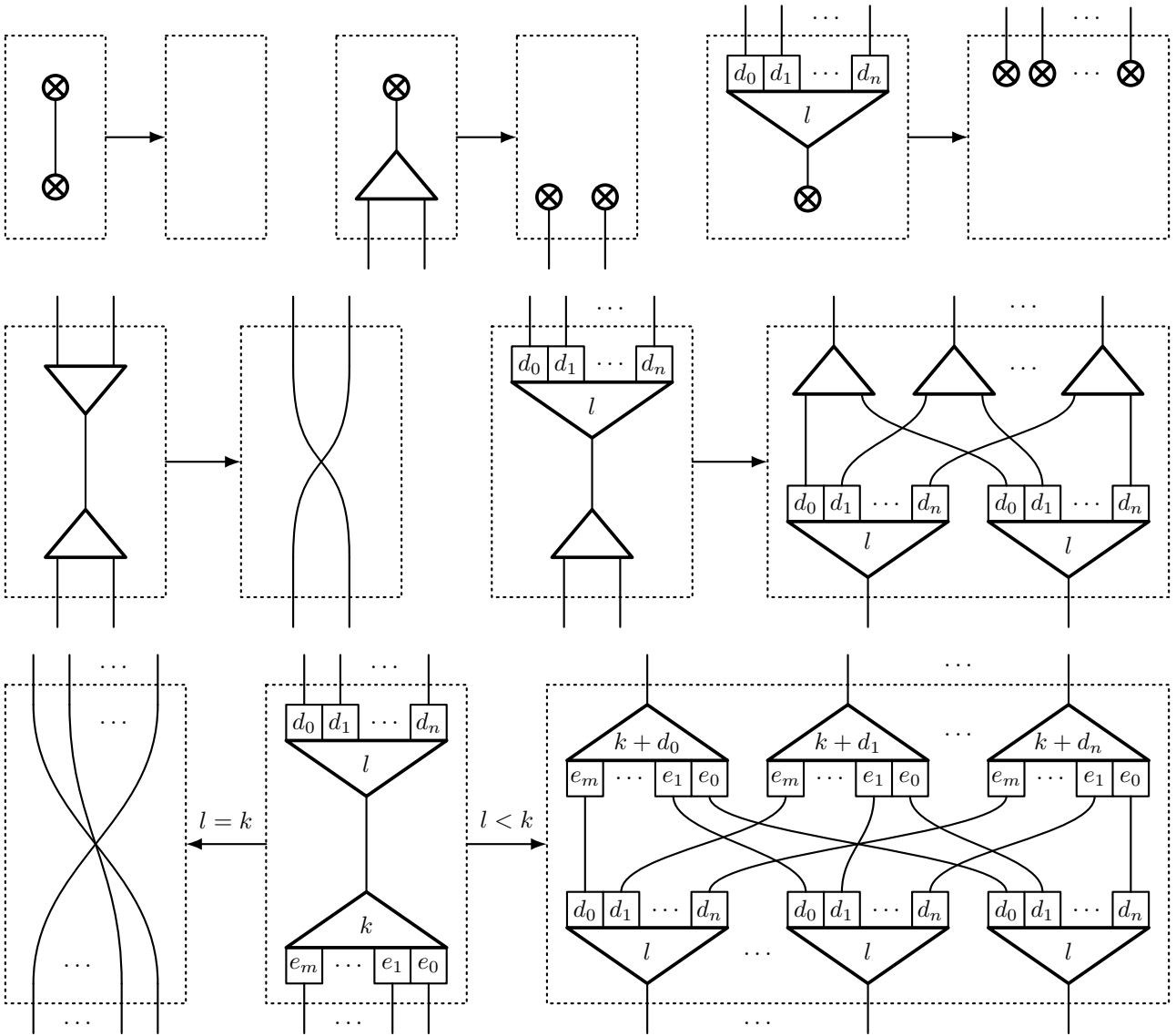

Fans are sufficient to express optimal parallel reduction in the \mbox{$\lambda L$-calculus}.
Erasers are needed to express optimal parallel reduction involving erasure.
Replicators are needed to express optimal parallel reduction involving sharing.
As such, the \mbox{$\Delta$-Nets} core interaction system decomposes perfectly into three overlapping subsystems, each analogous to a substructure \mbox{$\lambda$-calculus}. The full \mbox{$\Delta$-Nets} system may also be referred to as \mbox{$\Delta K$-Nets}. The four systems are:

\begin{itemize}
\item \textbf{\mbox{$\Delta L$-Nets}}: the \textit{linear} \mbox{$\Delta$-Nets} subsystem, which only uses fans, and expresses optimal parallel reduction in the \mbox{$\lambda L$-calculus}.
\item \textbf{\mbox{$\Delta A$-Nets}}: the \textit{affine} \mbox{$\Delta$-Nets} subsystem, which uses fans and erasers, and expresses optimal parallel reduction in the \mbox{$\lambda A$-calculus}.
\item \textbf{\mbox{$\Delta I$-Nets}}: the \textit{relevant} \mbox{$\Delta$-Nets} subsystem, which uses fans and replicators, and expresses optimal parallel reduction in the \mbox{$\lambda I$-calculus}.
\item \textbf{\mbox{$\Delta K$-Nets}}: the \textit{full} \mbox{$\Delta$-Nets} system, which uses fans, erasers and replicators, and expresses optimal parallel reduction in the \mbox{$\lambda K$-calculus}.
\end{itemize}

When equal agents interact, they \textit{annihilate} one another. In general, two replicators are equal if and only if they have the same level, number of auxiliary ports, and level deltas. However, when a \mbox{$\Delta$-net} is constructed from a \mbox{$\lambda$-term} via the translation method presented in the next section, interacting replicators that have the same level are guaranteed to be equal. As such, only replicator levels need to be compared for equality in practice. Equal-agent interactions are called \textit{annihilations}.

When distinct agents interact, each agent travels through and past the other and is potentially copied or erased in the process, depending on how many auxiliary ports the other agent has. Since an eraser has no auxiliary ports, it \textit{erases} every agent it interacts with. As such, distinct-agent interactions involving erasers are called \textit{erasures}. The remainder of distinct-agent interactions are called \textit{commutations}.
When a replicator interacts with a fan, the replicator travels through and out of the fan's two auxiliary ports, resulting in two exact copies of the replicator. Simultaneously, the fan travels through and out of all of the replicator's auxiliary ports, resulting in a fan for each replicator auxiliary port.

When two distinct replicators interact, the lower-level one replicates the higher-level one once for each of the lower-level one's auxiliary ports, while, simultaneously, the higher-level one duplicates the lower-level one once for each of the higher-level one's auxiliary ports. Note that whereas a \textit{duplication} produces exact copies, a \textit{replication} produces copies that may or may not be exact.
Replicators are so named because the copies they produce of other replicators are, by design, not necessarily exact. Each resulting replica of the higher-level replicator may have a different level, determined by the level delta associated to the auxiliary port of the lower-level replicator that it travels out of. The level of each resulting replica is the sum of the level of the original higher-level replicator and the appropriate level delta of the lower-level replicator. The \mbox{$\Delta$-Nets} interaction rules are illustrated in \mbox{Figure \ref{fig:lirules}}.

\section{From \mbox{$\lambda$-terms} to \mbox{$\Delta$-nets}}

\theoremstyle{plain}
\newtheorem*{definition}{Definition}
\newtheorem*{theorem}{Theorem}
\newtheorem*{corollary}{Corollary}

\begin{definition}
Let \mbox{$\Sigma = \{L,A,I,K\}$}. For all \mbox{$S \in \Sigma$}, let $\Lambda_S$ be the set of all \mbox{$\lambda S$-terms} and $\Delta_S$ be the set of all \mbox{$\Delta S$-nets}.
For all \mbox{$S \in \Sigma$}, there exists a bijection \mbox{$\phi_S: \Lambda_S \rightarrow \Delta_S^{c}$} which maps every \mbox{$\lambda S$-term} to a \emph{canonical} \mbox{$\Delta S$-net}.
\begin{align*}
\Delta_S^{c} &= \{ \phi_S(\lambda_S) \mid \forall \lambda_S \in \Lambda_S\} \tag{canonical \mbox{$\Delta S$-nets}} \\
\Delta_S^{p} &= \{ \delta_S^p \mid \forall \delta_S^c \in \Delta_S^{c},\ \delta_S^c \stackrel{\Delta^*}{\to} \delta_S^p \tag{proper \mbox{$\Delta S$-nets}} \}
\end{align*}
\[
  \Delta_S^{c} \subseteq \Delta_S^{p} \subset \Delta_S
\]
\end{definition}

\begin{figure}[H]
  \centering
  \begin{tikzpicture}[scale=0.58]

    \begin{scope}[xshift=8.75cm]
      \node at (1.1, 0.8 + 0.5 + 1.25) {$[\lambda x.M]_l,\ x \not\in FV(M)$};
      \draw[dotted] (-1.5, 0.8 + 0.5) -- (0.2+3.5, 0.8 + 0.5) -- (0.2+3.5,-4.5) -- (-1.5,-4.5) -- cycle;

      \fan{fan1}{0,0}{-1}{true};
      \node at (0,0.15) {$\lambda$};
      \eraser{era1}{-0.7,-0.4-0.3-1}{-1}{true};

      \draw (fan1-tip) -- ++(0, 1);
      \draw (fan1-left) -- (era1-tip);
      \draw (fan1-right) .. controls ++(0, -1) and ++(0, 1) .. (1.7, -1.4);

      \draw[dotted] (0.2, -1.4) -- (0.2+3, -1.4) -- (0.2+3,-4) -- (0.2,-4) -- cycle;
      \node at (1.7,-2.7) {$[M]_l$};

      \draw[dotted, line width=1mm, line cap=butt] (1.7, -4) -- (1.7, -4.5 -0.5);

    \end{scope}

    \begin{scope}
      \node[below left] at (6-0.3, 0.8 + 0.5 - 0.3) {$d_i = l_i - (l + 1)$ };
      \node at (2.25, 0.8 + 0.5 + 1.25) {$[\lambda x.M]_l,\ x \in FV(M)$};
      \draw[dotted] (-1.5, 0.8 + 0.5) -- (6, 0.8 + 0.5) -- (6,-7.5 -0.3 -0.5 -0.75) -- (-1.5,-7.5 -0.3 -0.5 -0.75) -- cycle;

      \fan{fan1}{0,0}{-1}{true};
      \node at (0,0.15) {$\lambda$};
      \replicator{rep1}{2,-6.55}{1}{1}{true};

      \draw (fan1-tip) -- ++(0, 1);
      \draw (fan1-left) -- (fan1-left |-, -7 -0.8) arc(180:270:0.5) -- (2-0.5, -7.5 - 0.8) arc(270:360:0.5) -- (rep1-tip);
      \draw (fan1-right) .. controls ++(0, -1) and ++(0, 1) .. (2.75, -1.4);

      \draw[dotted] (0, -1.4) -- (5.5, -1.4) -- (5.5,-4) -- (0,-4) -- cycle;
      \node at (2.75,-2.7) {$[M]_l$};

      \draw[dotted, line width=1mm, line cap=butt] (2+2+0.75, -4) -- (2+2+0.75, -7.5 -0.3 -0.5 -0.5 -0.75);

      \draw (rep1-d0w) -- ++(0, 1.25) node[midway, left] {$l_0$};
      \draw (rep1-d1w) -- ++(0, 1.25) node[midway, left] {$l_1$};
      \draw (rep1-dnw) -- ++(0, 1.25) node[midway, left] {$l_n$};
      \node at (rep1-d0) {$d_0$};
      \node at (rep1-d1) {$d_1$};
      \node at (rep1-dn) {$d_n$};
      \node[anchor=base] at (2,-6.825) {$l + 1$};
    \end{scope}

    \begin{scope}[yshift=-13.25cm]
      \node at (2.25, 0.8 + 1 + 0.9)  {$[M N]_l$};
      \draw[dotted] (-1.5, 0.8 + 1) -- (6, 0.8 + 1) -- (6,-4) -- (-1.5,-4) -- cycle;

      \fan{fan1}{0.5,0.4}{1}{true};
      \node at (0.5, 0.35) {$@$};

      \draw (fan1-left) -- ++(0, 1.5);
      \draw (fan1-tip) -- ++(0, -0.5);
      \draw (fan1-right) arc(180:90:0.5) -- (3.5,0.4+0.4+0.5) arc(90:0:0.5) -- (4, -1.4+0.5);

      \draw[dotted] (-1, -1.4+0.5) -- (2, -1.4+0.5) -- (2,-4+0.5) -- (-1,-4+0.5) -- cycle;
      \node at (0.5,-2.7+0.5) {$[M]_l$};

      \draw[dotted] (-1+3.5, -1.4+0.5) -- (2+3.5, -1.4+0.5) -- (2+3.5,-4+0.5) -- (-1+3.5,-4+0.5) -- cycle;
      \node at (4,-2.7+0.5) {$[N]_{l+1}$};

      \draw[dotted, line width=1mm, line cap=butt] (0.5, -3.5) -- ++(0, -1);
      \draw[dotted, line width=1mm, line cap=butt] (4, -3.5) -- ++(0, -1);
    \end{scope}

    \begin{scope}[xshift=8.75cm, yshift=-8cm]
      \node at (1.1, 1.25) {$[x]_l,\ x \not\in FV(*)$};
      \draw[dotted] (-1.5, 0) -- (3.7, 0) -- (3.7,-2) -- (-1.5,-2) -- cycle;
      \draw (1.05, 0.5) -- (1.05, -2);
      \draw (1.05, -2) -- (1.05,-2-0.5-0.75) node[midway, left] {$l$};
    \end{scope}

    \begin{scope}[xshift=8.75cm, yshift=-14.25cm]
      \node at (1.1, 1.25) {$[x]_l,\ x \in FV(*)$};
      \draw[dotted] (-1.5, 0) -- (3.7, 0) -- (3.7,-3) -- (-1.5,-3) -- cycle;
      \draw (1.05, 0.5) -- (1.05, -1);
      \node at (1.05, -1.5) {$x$};
    \end{scope}

  \end{tikzpicture}

  \caption{The inductive definition of $\phi_K$, a bijection which translates \mbox{$\lambda K$-terms} into canonical \mbox{$\Delta K$-nets}.}
  \label{fig:ltrules}
\end{figure}
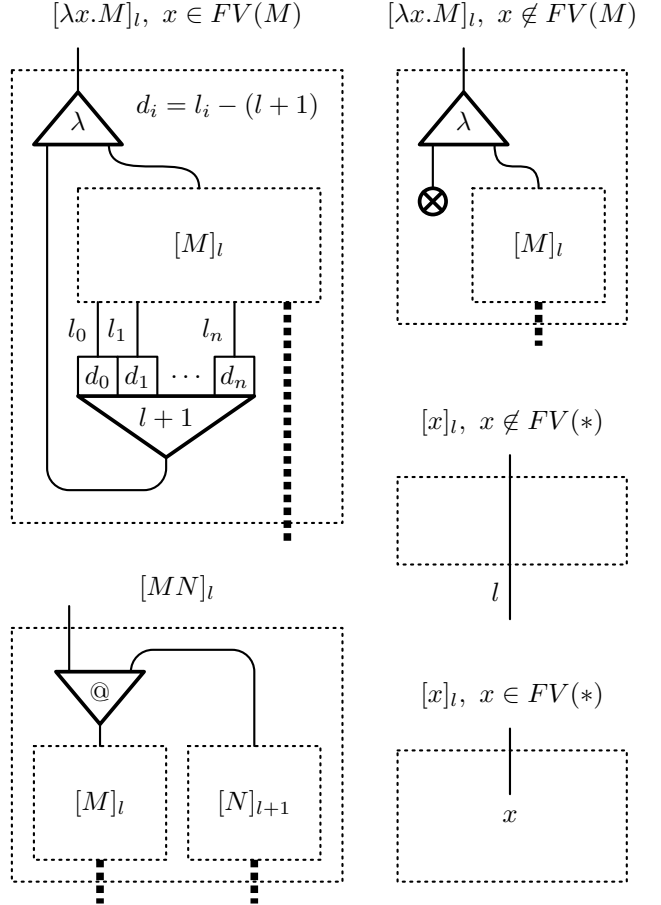

The bijections $\phi_S,\ \forall S \in \Sigma$, are defined inductively, with the rules for the general $\phi_K$ case illustrated in \mbox{Figure \ref{fig:ltrules}}.
In the description that follows, the minor differences in \mbox{$\phi_L$}, \mbox{$\phi_A$}, and \mbox{$\phi_I$} are noted where appropriate.
Each outer dashed rectangle in \mbox{Figure \ref{fig:ltrules}} contains a \mbox{$\Delta$-net} fragment that represents the \mbox{$\lambda$-term} specified above it.
Each inner dashed rectangle is a slot for the \mbox{$\Delta$-net} that represents the inner term.
Each thick vertical dashed line represents a non-negative integer number of parallel wires (including zero). Every dashed rectangle, outer or inner, has the same interface---a single wire entering at the top and a non-negative integer number of wires leaving at the bottom.
Each \mbox{$\lambda$-term} has a subscript associated with it, which represents the \textit{level} of that term.
The level of the outermost term is set to zero, which inductively sets all other levels. The level of an application's argument is one greater than that of the application itself, and the level of a replicator is one greater than that of its associated abstraction.
These levels are ultimately used to determine the level and level deltas of replicators.

There are two \mbox{$\Delta$-net} fragments for variables, one for variables that are free in the outermost \mbox{$\lambda$-term} (\mbox{$x \in FV(*)$}), and one for variables that are not (\mbox{$x \not\in FV(*)$}), i.e., that are bound by some abstraction.
The free-variable fragment contains a single-port (non-agent) node, which is represented by the name of the associated free variable in the \mbox{$\lambda$-term}.
The bound-variable fragment is just a vertical wire.
An instance of the bound-variable fragment which represents the $i$th occurrance of a bound variable in the associated \mbox{$\lambda$-term} has its bottom wire endpoint connected to the $i$th auxiliary port of the replicator that shares that variable.
The bottom wire endpoint of each bound-variable fragment has a level associated with it, which is equal to the level of the variable.
These wire endpoint levels are referenced in the fragment that features a replicator, in the definition of the level deltas $d_i$. The level delta associated to an auxiliary port of a replicator is equal to the level of the wire connected to that auxiliary port minus the level of the replicator.

There are also two \mbox{$\Delta$-net} fragments for abstractions, one for those which use their bound variables and one for those which don't.
In both, the abstraction itself is represented by a fan pointing up, labelled with a $\lambda$\footnote{\label{visualaid}Fan labels are just a visual aid and do not affect how fans interact with other agents.}.
The principal port of the fan is the parent port of the abstraction.
The first auxiliary port of the fan is a child port, connected to the \mbox{$\Delta$-net} that represents the body of the abstraction. The second auxiliary port represents the variable of the abstraction, and since it is connected to the parent of the variable, it is another parent port.
If an abstraction doesn't use its bound variable, the fan's second auxiliary port is connected to an eraser.
The fragment with this eraser is only needed when translating expressions from \mbox{$\lambda$-calculi} with erasure, and, as such, it is exclusively part of $\phi_A$ and $\phi_K$.
When translating expressions from {$\lambda$-calculi} with sharing, if an abstraction uses its bound variable, then the fan's second auxiliary port is generally connected to the principal port of a replicator, which shares the abstraction's variable among the bound-variable fragments that represent its various occurances in the \mbox{$\lambda$-term}. However, in a canonical \mbox{$\Delta$-net}, a replicator with a single auxiliary port and a level delta of zero, regardless of the replicator's level, is equivalent to a wire. Therefore such replicators are never created in the first place, and the appropriate bound-variable fragment's bottom wire is connected directly to the abstraction fan's second auxiliary port instead.
In \mbox{$\phi_L$} and \mbox{$\phi_A$}, which translate \mbox{$\lambda$-terms} without sharing, the fragment with the replicator is modified such that the replicator is substituted by a single wire in the same way.

Finally, there is a fragment for applications, in which the application itself is represented by a fan pointing down, labelled with an $@$\footref{visualaid}.
The first auxiliary port of the fan is the parent port of the application.
The principal port of the fan is a child port, connected to the \mbox{$\Delta$-net} that represents the function \mbox{$\lambda$-term}. The second auxiliary port is another child port, connected to the \mbox{$\Delta$-net} that represents the argument \mbox{$\lambda$-term}.
In addition to the fragments shown in \mbox{Figure \ref{fig:ltrules}}, every canonical \mbox{$\Delta$-net} has a single \textit{root} (non-agent) node, represented by a small circle, connected to the top of the outermost \mbox{$\Delta$-net} fragment.
Together, the root node and the free variable nodes constitute the interface of a canonical \mbox{$\Delta$-net}.

Note that both applications and abstractions are represented by fans, and \mbox{$\beta$-reduction} is expressed through fan annihilation.
Whether a specific fan in a \mbox{$\Delta$-net} represents an abstraction or an application can always be determined by inspecting the net (without fan labels) and tracing paths from the root node.
While abstraction fans have two parent ports and one child port, application fans have one parent port and two child ports.
Every port of any node (including non-agent nodes) in a proper \mbox{$\Delta$-net} is either a child port or a parent port, and every wire connects a child port with a parent port.
Although this duality has no direct bearing on the interaction process, it is clearly present, and it is instrumental in some discussions.

All replicators in a canonical \mbox{$\Delta$-net} are unpaired fan-ins: each auxiliary port is a parent port and the principal port is a child port.
However, during reduction, fan-out replicators may be produced.
In a fan-out replicator, the principal port is a parent port and each auxiliary port is a child port.
Every commutation between a fan and a replicator (either a fan-in or a fan-out) always produces a fan-in and a fan-out.
While every fan-out is paired with at least one upstream fan-in, the converse is not true: fan-ins may or may not be paired. Locally determining this pairing efficiently is the purpose of the level delta system.

In some ways, a fan is just a replicator with two auxiliary ports, zero level, and zero deltas. If replicators were allowed to have zero auxiliary ports, then an eraser could also just be a replicator with no auxiliary ports and zero level (or no level). It's tempting to attempt to consolidate the three agents into one---the \mbox{$\Delta$-agent}, if you will. However, fans and replicators have a critical distinction in the types of their auxiliary ports: a replicator's auxiliary ports are either all parent ports or all child ports, while a fan has one of each. This affects how replicator pairedness is tracked: after fan replication the resulting replicators become paired whereas in replicator replication they keep their original status.

In standard \mbox{$\Delta$-Nets}, replicators have \emph{absolute levels}, i.e., the initial level of each replicator is exactly determined by how many times a path from the root to the replicator's principal port traverses fans out of their second auxiliary ports.
There exists a dual formulation of the \mbox{$\Delta$-Nets} system to this where replicators have \emph{relative} levels---all replicators start with level zero, and a replicator's level gets incremented when it traverses out of the second auxiliary port of a fan.

\section{From \mbox{$\Delta$-nets} back to \mbox{$\lambda$-terms}}

The only interaction rule in \mbox{$\Delta L$-Nets} is fan annihilation, which expresses \mbox{$\beta$-reduction}.
Therefore if $n$ \mbox{$\beta$-reductions} normalize a \mbox{$\lambda L$-term} $t$, then $n$ interactions normalize the \mbox{$\lambda L$-net} \mbox{$\phi_L(t)$}.
As a result, in the \mbox{$\Delta L$-Nets} system, all proper nets are canonical, and fan annihilations can be applied in any order, with perfect confluence. Compared to \mbox{$\Delta L$-Nets}, the other \mbox{$\Delta$-Nets} systems have additional interaction rules, which don't have \mbox{$\lambda$-calculus} analogues.

In \mbox{$\Delta $-Nets} systems with erasure, applying an abstraction which doesn't use its bound variable results in an eraser becoming connected to a parent port.
Such an eraser could erase abstraction fans and fan-out replicators, but if it reaches the parent port of an application fan, for example, which is an auxiliary port, the erasure process would cease.
As a result, absent additional rules, irreducible subnets would be produced which don't have a \mbox{$\lambda$-calculus} analogue.
As an extreme example, applying an abstraction which doesn't use its bound variable to an argument which only uses globally-free variables produces a subnet which is disjointed from the root.
In order to eliminate all such subnets a final \emph{canonicalization} reduction step is introduced in \mbox{$\Delta$-Nets} systems with erasure: all parent-child wires starting from the root are traversed and nodes are marked. All non-marked nodes are then erased, and wires that were connected to these nodes are instead connected to erasers.
This final canonicalization erasure step dispenses with the need to apply erasure and eraser annihilation rules.
In fact, this step can be applied at any point during reduction in order to reduce the net size, effectively trading computation (time) for memory (space). In order to keep memory usage to a minimum, this step should be applied after every application of an abstraction which doesn't use its bound variable.

In order to ensure that no reduction operations are applied in a subnet that is later going to be erased, a sequential leftmost-outermost reduction order needs to be followed.
Therefore, in the \mbox{$\Delta A$-Nets} system, fan annihilations are applied in leftmost-outermost order, with the final erasure canonicalization step ensuring perfect confluence, and producing a normal canonical \mbox{$\Delta A$-net}.

\begin{figure*}[!t]
  \centering
  \begin{tikzpicture}[scale=0.58]

    \begin{scope}
      \replicator{origdup}{0.5,3.2-0.5}{1}{1}{true}
      \replicator{sup}{4,4}{-1}{-1}{false}
      \fan{fan1}{0,0}{1}{false};
      \fan{fan2}{2,-2}{1}{true};
      \fan{fan3}{7,0}{1}{false};
      \replicator{dup}{3,-3}{1}{1}{false}

      \node at ([shift={(0,1.95+0.5)}]origdup-dots) {$\ldots$};

      \node at (0.5,3.2-0.5-0.2) {$l$};
      \node at (origdup-d0) {$d_0$};
      \node at (origdup-d1) {$d_1$};
      \node at (origdup-dn) {$d_n$};

      \draw (sup-tip) -- ++(0, 1.25);
      \draw (fan2-tip) -- ++(0, -1.675 -0.75);
      \draw (fan2-left) .. controls ++(0, 1.5) and ++(0, -1.5) .. (origdup-tip);
      \draw (origdup-d0w) -- ++(0, 2.25);
      \draw (origdup-d1w) -- ++(0, 2.25);
      \draw (origdup-dnw) -- ++(0, 2.25);
      \draw (fan2-right) .. controls ++(0, 1.5) and ++(0, -4) .. (sup-tip);

      \draw[dotted] (-2, 5.5) -- (4+0.45+0.5, 5.5) -- (4+0.45+0.5, -4.5) -- (-2, -4.5) -- cycle;
    \end{scope}

    \draw[->] (4+0.45+0.5, 0.5) -- (4+0.45+0.5 + 1.25, 0.5);

    \begin{scope}[xshift=6.95cm + 1.25cm]
      \replicator{origdup}{0.5,4.2-0.5}{1}{1}{false}
      \replicator{sup}{6,4}{-1}{-1}{true}
      \fan{fan1}{0,0}{1}{true};
      \fan{fan2}{3,0}{1}{true};
      \fan{fan3}{7,0}{1}{true};
      \replicator{dup}{3,-3}{1}{1}{true}

      \node at (origdup-dots) {$\ldots$};
      \node at ([shift={(0,0.95+0.5)}]origdup-dots) {$\ldots$};
      \node at (5, -0.25) {$\ldots$};

      \node at (3,-3-0.2) {$l$};
      \node at (6,4+0.2) {$l$};
      \node at (dup-d0) {$d_0$};
      \node at (dup-d1) {$d_1$};
      \node at (dup-dn) {$d_n$};
      \node at (sup-d0) {$d_0$};
      \node at (sup-d1) {$d_1$};
      \node at (sup-dn) {$d_n$};

      \draw (sup-tip) -- ++(0, 1.25);
      \draw (dup-tip) -- ++(0, -1.25);
      \draw (fan1-tip) .. controls ++(0, -0.8) and ++(0, 0.8) .. (dup-d0w);
      \draw (fan2-tip) .. controls ++(0, -0.6) and ++(0, 0.6) .. (dup-d1w);
      \draw (fan3-tip) .. controls ++(0, -0.8) and ++(0, 0.8) .. (dup-dnw);
      \draw (fan1-left) .. controls ++(0, 2) and ++(0, -3) .. (origdup-d0w);
      \draw (fan2-left) .. controls ++(0, 2) and ++(0, -4) .. (origdup-d1w);
      \draw (fan3-left) .. controls ++(0, 2) and ++(0, -5) .. (origdup-dnw);
      \draw (origdup-d0w) -- ++(0, 1.25);
      \draw (origdup-d1w) -- ++(0, 1.25);
      \draw (origdup-dnw) -- ++(0, 1.25);
      \draw (fan1-right) .. controls ++(0, 2) and ++(0, -3) .. (sup-d0w);
      \draw (fan2-right) .. controls ++(0, 2) and ++(0, -1) .. (sup-d1w);
      \draw (fan3-right) .. controls ++(0, 2) and ++(0, -1) .. (sup-dnw);

      \draw[dotted] (-2, 5.5) -- (8.5, 5.5) -- (8.5, -4.5) -- (-2, -4.5) -- cycle;
    \end{scope}

  \end{tikzpicture}
  \caption{The aux fan replication canonicalization rule in \mbox{$\Delta I$-} and \mbox{$\Delta K$}-Nets.}
  \label{fig:ipsi}
\end{figure*}
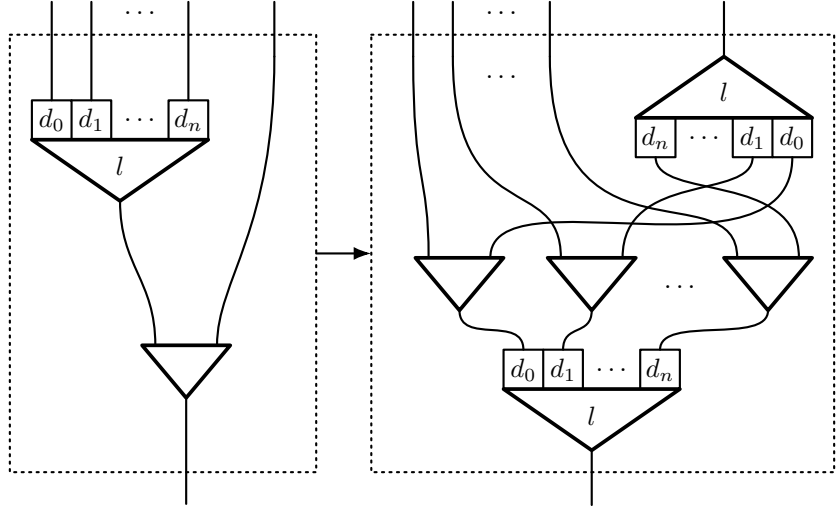

During reduction of \mbox{$\Delta I$-} and \mbox{$\Delta K$-Nets}, replicators can combine into \textit{replicator trees}. A replicator tree is a subnet containing only replicator agents such that each one's principal port is connected to an auxiliary port of another, except for the tree's \textit{root} replicator.
In general, replicators that  belong to the same tree cannot be merged during reduction because they may be paired with other replicators elsewhere. However, if consecutive replicators in a replicator tree are known to be unpaired, they can be safely merged together.
\emph{Unpaired replicator merging} is a canonicalization rule present in both \mbox{$\Delta I$-} and \mbox{$\Delta K$-Nets}.
Additionally, in the \mbox{$\Delta K$-Nets} system, it is possible to eliminate unpaired replicators' auxiliary ports which are connected to erasers---a canonicalization rule called \emph{unpaired replicator decay}.
As part of unpaired replicator decay, if an unpaired replicator is left with a single auxiliary port with a level delta of zero, it is replaced by a wire, as it is equivalent to one.
This canonicalization rule is applied to all unpaired replicators as part of the erasure canonicalization step. It can also be lazily applied to the involved unpaired replicator, immediately before the fan replication, replicator replication, and aux fan replication rules.

Replicators start as \emph{unpaired}, and this status is propagated across interactions.
When an unpaired replicator interacts with a fan, the status of both resulting replicators changes to \emph{unknown}.
If an unpaired replicator (A) is connected to a consecutive replicator (B) of unknown status via an auxiliary port, and a certain local constraint is met, then the consecutive replicator can be determined to be unpaired, and the two can then be merged.
The constraint is met when the second replicator's level is greater than or equal to the first replicator's level, but no greater than the first replicator's level plus the level delta of the auxiliary port that connects them: \mbox{$0 \le l_B - l_A \le d$}. Under this constraint, no replicator is able to interact with the second replicator before the first replicator is annihilated. Since the first replicator is unpaired, it can never be annihilated, and the second one must be unpaired as well.

It is possible to limit replicators to have at most two auxiliary ports, or even exactly two auxiliary ports. This imposes the smallest possible upper bound on interaction rule complexity and agent size, making the total number of interactions an effective measure of time complexity, and the total number of agents an effective measure of space complexity. A tree of such replicators can stand in for any replicator with any number of auxiliary ports.

Replicator merging is a canonicalization rule and not an interaction rule because it involves two agents that are connected via ports that aren't both principal.
Merging replicators as early as possible reduces the total number of reductions and the total number of agents, improving space and time efficiency.
The reduction order which guarantees that replicator merges happen as early as possible, minimizing the total number of reductions, is a sequential leftmost-outermost order---the optimal reduction order for \mbox{$\Delta A$-}, \mbox{$\Delta I$-} and \mbox{$\Delta K$-Nets}. In fact, any reducible pair that eventually reaches the leftmost-outermost position unchanged can be reduced at any time.
For example, all potential annihilations in the spine that don't involve unpaired replicators can be applied in any order, because they will eventually reach the top of the spine unchanged. On the other hand, a potential commutation that involves an unpaired replicator may not reach the top of the spine unchanged, as the involved replicator may be merged with another beforehand, rendering its early application suboptimal.

To see how leftmost-outermost reduction is optimal in \mbox{$\Delta I$-} and \mbox{$\Delta K$-Nets}, observe that no commutation involving an unpaired replicator can be applied before that replicator is merged---if merging it is at all possible. Take a leftmost-outermost interaction between two replicators. The fan-out replicator is necessarily paired.
If the fan-in replicator is unpaired and can eventually be merged, then the replicator merging---or any intermediate reductions leading to it---would occur higher in the spine and would be applied first.
This reasoning holds even in the presence of loops.
In \mbox{$\Delta K$-Nets}, the leftmost-outermost order is critical not only to achieve optimality but also to ensure that all nets associated with normalizing \mbox{$\lambda$-terms} normalize.
As for the remaining rules, the \mbox{$\Delta$-Nets} systems inherit their optimality guarantee from the interaction nets paradigm.

Additionally, in \mbox{$\Delta I$-} and \mbox{$\Delta K$-Nets}, the reduction process needs to be split in two phases. In the first phase the core interaction rules and unpaired replicator merging are applied in leftmost-outermost order until no further reduction can be applied. The reduction process then switches to the second phase, in which the \emph{aux fan replication} rule (illustrated in \mbox{Figure \ref{fig:ipsi}}) replaces the core fan replication rule.
This second phase can be alternatively understood as modifying all fans such that the first auxiliary port becomes the principal port. This process transforms the sharing structures so that all fan-out replicators are eliminated, all appropriate subnets are replicated, and all fan-in replicators accumulate at the variable port of abstraction fans.
The result is a canonical \mbox{$\Delta I$-net} or, after the final erasure canonicalization step, a canonical \mbox{$\Delta K$-net}.
Since all normal \mbox{$\Delta$-nets} are canonical, the \mbox{$\Delta$-Nets} systems are all Church--Rosser confluent.

\vspace{0.3cm}
\begin{definition}
For all \mbox{$S \in \Sigma$} and for all \mbox{$\lambda_S \in \Lambda_S$}, if $\lambda_S$ is normalizing, \mbox{$\lambda_S$} \mbox{$\beta$-reduces} to \mbox{$\phi_S^{-1}(\Omega_S(\phi_S(\lambda_S)))$} where \mbox{$\Omega_S : \Delta_S^p \rightarrow \Delta_S^{c}$} reduces a proper \mbox{$\Delta S$-net} through interaction and canonicalization rules as defined in this section until it is normal and canonical.
\end{definition}

\begin{theorem}
Since \mbox{$\phi_S^{-1} \circ \Omega_S$} is idempotent \mbox{$\forall S \in \Sigma$}, the set of all \mbox{$\lambda S$-terms}, \mbox{$\Lambda_S$}, is a projection of proper \mbox{$\Delta S$-nets}:
\vspace{-0.05cm}
\begin{align*}
  \Lambda_S = \{\phi_S^{-1}(\Omega_S(\delta_S^p)) \mid \forall \delta_S^p \in \Delta_S^{p} \},\  \forall S \in \Sigma
\end{align*}
Moreover, since \mbox{$\Lambda_S$} is closed under \mbox{$\beta$-reduction} and \mbox{$\Delta_S$} is closed under \mbox{$\Delta_S$-interactions} and canonicalizations, the \mbox{$\lambda S$-calculus} can be interpreted as a projection of \mbox{$\Delta S$-Nets}, \mbox{$\forall S \in \Sigma$}.
\end{theorem}

\vspace{-0.1cm}
\section{Conclusion}

The existence of nonlinear proper \mbox{$\Delta$-nets} which are not canonical reflects the additional degrees of freedom that interior sharing introduces, which are not present in the \mbox{$\lambda$-calculus}.
A given \mbox{$\lambda I$-term} can potentially be represented by many different proper \mbox{$\Delta I$-nets}, which differ only with respect to their sharing structure.
As an example, take a proper \mbox{$\lambda I$-net} $n$, with \mbox{$\phi_I^{-1}(\Omega_I(n)) = M M$}, for some \mbox{$\lambda I$-term} $M$.
The \mbox{$\lambda I$-net} $n$ may have two distinct but equal subnets that each represent $M$, or it may have a single such subnet which is shared among the two occurrances in the application.
In fact, the sharing structure of $n$ could be arbitrarily complex.
These additional degrees of freedom allow \mbox{$\Delta$-Nets} to realize optimal reduction in the manner envisioned by Lévy, i.e., no reduction operation is applied which is rendered unneccessary later, and no reduction operation which is necessary is applied more than once.

\end{multicols}
\end{spacing}
\end{document}